# Probing the carrier dynamics of polymer composites with single and hybrid carbon nanotube fillers for improved thermoelectric performance

Ioannis Konidakis*[1], Beate Krause[2], Gyu-Hyeon Park[3], Nithin Pulumati[3], Heiko Reith[3], Petra Pötschke[2], Emmanuel Stratakis*[1]

[1]Institute of Electronic Structure and Laser (IESL), Foundation for Research and Technology-Hellas (FORTH), 70013, Heraklion-Crete, Greece.

[2]Leibniz-Institut für Polymerforschung Dresden e.V. (IPF), Hohe Str. 6, 01069 Dresden, Germany.

[3]Leibniz-Institut für Festkörper- und Werkstoffforschung Dresden e.V. (IFW), Helmholtzstr. 20, 01069 Dresden, Germany.

**\*Corresponding authors:**

Phone: +30-2810-392944, ikonid@iesl.forth.gr

Phone: +30-2810-391274, stratak@iesl.forth.gr

**Abstract**

The incorporation of carbon nanotubes (CNTs) within polymer hosts offers a great platform for the development of advanced thermoelectric (TE) composite materials. Over the years, several CNT/polymer composite formulations have been investigated on an effort to maximize the TE performance. Meanwhile, several studies focused on the decay dynamics of the charged excitons within CNTs itself and therefrom derived structures, aiming to investigate the lifetimes and the corresponding recombination processes of free charge carriers. The latter physical phenomena play a crucial role in the performance of various types of energy converting and scavenging materials. Nevertheless, up to this date, there is no systematic study on the combination of TE parameters and the critical charge carrier dynamics within CNT containing TE polymer composites. Herein, a variety of composites with single and hybrid CNT fillers based on polycarbonate (PC) and polyether ether ketone (PEEK) polymer matrices were prepared by melt-mixing in small scale. At the same loading, the addition of single fillers in PC results in higher Seebeck coefficients and similar conductivities when compared to the use of hybrid filler systems. In contrast, with hybrid filler systems in PEEK composites, higher power factors could be reached than in single filler composites. Moreover, the PC-based composites are studied using ultrafast laser time-resolved transient absorption spectroscopy (TAS), for the investigation of the exciton lifetimes and the physical origins of free charge carrier transport within the TE films. The findings of this study reveal interesting links between the TE parameters and the obtained charge carrier dynamics. Namely it is found that the Seebeck coefficient of the composites relates directly to the exciton lifetimes, whereas the volume conductivity is independent of the exciton lifetimes and is determined mainly by the average number and the mobility of the free charged electrons at the higher energy states.

## 1. Introduction

Over the years, carbon nanotubes (CNTs) have attracted significant scientific interest due to their fascinating optical, photophysical, and electrical properties, directed towards various photonic and optoelectronic[1-3] as well as energy harvesting and storage applications.[4,5] With regard to the latter, the incorporation of CNTs into suitable intrinsically conductive (ICP) or nonconductive, especially thermoplastic, polymers offers a promising approach on creating advanced thermoelectric (TE) nanocomposite materials for converting waste heat to electrical energy.[6-8] The advantages of thermoplastic polymer-based TE materials over typically used metal oxides are not only the replacement of rare earth elements, their better availability and cost efficiency, but also their ease of processing, mechanical flexibility, low density and low thermal conductivity. In the context of larger scale industrial applications, TE composite materials based on insulating thermoplastic polymers and CNTs seem to be worth to be developed, despite the expected lower efficiency compared to ICPs or ICP-based composites.

Along these lines, single-walled CNTs (SWCNTs), multi-walled CNTs (MWCNTs) and boron- or nitrogen doped CNTs have been employed in an attempt to maximize the performance of such developed melt-mixed TE polymer composite films.[9-10] Notably, nearly all industrial synthesized SWCNTs contain a mixture of semi-conductive and metallic nanotubes. Statistically, in defect-free SWCNT materials about 1/3 are metallic and 2/3 are semiconducting. For the used type of SWCNT the producer does not give a ratio. However, the results reflect the typical behavior of an industrial (mixed chirality) SWCNT product. As no single chirality tubes can be separated or are available, the behavior of them cannot be studied practically. Also, industrial MWCNTs contain several (typically 8-12) graphene layers with different chirality so that on individual nanotubes as well conductive as semiconductive behavior was measured which implies that also the TE properties of different MWCNTs may differ. In addition, the polymer matrix also influences the TE performance as for instance by doping the CNTs, although the TE effect is solely due to the incorporated CNT network. This mutual influence means that, for example, the incorporation of p-type SWCNTs can lead to both p-type and n-type polymer composites.[10] Depending on the aimed application of the composites a variety of thermoplastic polymer matrices were described in the literature to generate such TE materials, as for instance thermoplastic polyurethane (TPU),[11] polycarbonate (PC),[12] polypropylene (PP),[13] polyvinylidene fluoride (PVDF),[8] polyamide (PA),[10] polybutylene terephthalate (PBT),[10] acrylonitrile butadiene styrene (ABS),[10] and polyether ether ketone (PEEK).[14] Notably, other types of materials apart from CNTs have been also employed over the years for the development of advanced TE architectures of superior performance, as for instance $Bi_2Te_3$ and $Bi_2GeTe_4$-based configurations,[15,16] while the incorporation of suitable dopands like dimethyl sulfoxide is also a highly promising approach.[17,18]

Meanwhile, during the last decade, the transient exciton dynamics of CNTs has been the subject of numerous studies.[19-24] This is because the free-charge carrier (electrons and holes) lifetime and recombination and transfer processes strongly determine the optical and electronic features of CNT-based materials. Thus, studying and shedding light on the physical mechanisms behind these ultrafast procedures is expected to pave the way towards the design also of polymer-based composites with advanced TE properties. Namely, ultrafast laser time-resolved transient absorption spectroscopy (TAS) offers a great tool for probing the picosecond regime processes of exciton creation and free-charge carrier transfer within SWCNTs.[19-24] Based on the operation principle of pump-probe spectroscopy, a light source is used to photoexcite electrons, while the corresponding decay dynamics of the consequent relaxation process is monitored in terms of optical absorption within various time delays. Remarkably, it was proven that in SWCNTs that pose an excellent approximation of one dimension quantum confinement, charge carriers are formed instantaneously upon photoexcitation, i.e. within 50 fs.[21] In addition, following the typical electron excitations to higher energy states within SWCNTs, it was also revealed that the excitons may diffuse plausibly for the formation of trions.[22] The latter consist of an exciton bonded to another positively charged hole located in the network of SWCNTs. Typically, the revealed recombination periods from time-resolved spectroscopy studies for all types of excitonic processes within SWCNTs are reported to be of several ps. More importantly, in energy converting devices such as organic photovoltaics,[25,26] and perovskite solar cells,[27,28] the correlation between the obtained exciton dynamics and the power conversion efficiency is striking. In particular, it has been found that longer exciton lifetimes and slower recombination rates are indicative of enhanced device efficiency and stability.[25-28] However, to date, although these are considered critical, studies and especially a systematic correlation of the CNT filler type and the concentration effect on the transient exciton dynamics and consequently on the TE performance of polymer based composite systems are lacking.

In this study, a series of TE composites based on single- and hybrid-CNT fillers in PC and PEEK were prepared by melt-mixing. The therefrom prepared thin films are characterized by microscopy techniques, while their TE power conversion performance is determined in terms of Seebeck coefficient, S, and electrical volume conductivity, σ. Moreover, the mobility of the charge carriers in both single- and hybrid-CNT composite systems with PC was investigated by Hall measurements. Ultrafast laser time-resolved TAS was employed to thoroughly investigate the exciton dynamics and corresponding charge carrier lifetimes within the developed PC/CNT composites. The findings of this study shed light on the physical origins of the TE conversion efficiency by providing evidence of how it is related with the excitonic characteristics of the incorporated CNTs. Namely, it is revealed that the Seebeck coefficient of both single- and hybrid-filler composites exhibits a direct correlation with the exciton lifetime of the CNTs, while being nearly independent of the CNT concentration within the host polymer.

## 2. Experimental

### 2.1 Materials

As polymer matrices polycarbonate (Makrolon® 2600, Bayer MaterialScience, Germany) with a density of 1.2 g/cm$^3$ and a melt volume-flow rate (MVR) of 12 cm$^3$/10 min (300 °C/1.2 kg) as well as polyether ether ketone (Vestakeep 1000P, Evonik, Germany), a material with a density of 1.3 g/cm$^3$ and a MVR of 140 cm$^3$/10 min (380°C/5 kg), were selected. The two polymers were selected to test whether the effects found on the TE properties depend on the type of the polymer matrix, whereby PC represents an amorphous and PEEK a partially crystalline polymer. Three kinds of commercially available CNTs were chosen for this study, namely SWCNT Tuball™ (carbon purity 75%, OCSiAl S.a.r.l., Luxembourg, Luxembourg, abbreviated Tuball),[29] MWCNT NC7000™ (carbon purity 90%, Nanocyl, S.A., Sambreville, Belgium, abbreviated NC7000),[30] and branched MWCNT Athlos™ 200 (Cabot Corp., USA; previously CNS-PEG, Applied NanoStructured Solutions LLC, Baltimore, MD, US, abbreviated CNS-PEG).[31] RAMAN spectra and D/G intensity ratios of the different CNT materials are presented in Figure S1.

### 2.2 Methods

Melt mixing of the composites was performed in a small-scale conical twin-screw micro compounder (Xplore Instruments BV, Sittard, The Netherlands) having a volume of 15 cm$^3$. The composites were prepared at 280°C (PC) or 360 °C (PEEK) with a rotation speed of 250 rpm for 5 min. The polymer granules (PC) or polymer powder (PEEK) and CNTs were alternately filled into the main hopper of the compounder. The extruded strands were compressing molded at the melt mixing temperature for 1 min into plates (60 mm diameter, 0.3 mm or 10 µm thickness) using the hot press PW40EH (Paul-Otto Weber GmbH, Remshalden, Germany). Strips cut from such plates were used as films for the measurements of the TE properties. All CNT contents are given in wt% based on the polymer weight.

The morphological characterization of the composites was performed using scanning electron microscopy (SEM) by means of a Carl Zeiss Ultra plus microscope combined with a SE2 detector. Before imaging, the composite strands were cryo-fractured in liquid nitrogen, and the surfaces were coated with 3 nm platinum. The macro dispersion of the CNTs in the polymer matrices was characterized by transmission light microscopy (TLM). The extruded strands were first cut into thin sections with a thickness of 5 µm using a Leica RM2265 microtome at room temperature and then fixed on glass slides using the embedment medium Aquatex®. The TLM investigations were performed using an Olympus BX53M microscope combined with an Olympus DP74 camera.

TE measurements were performed with the home-built TEG facility at IPF at 40 °C with temperature differences between the two copper electrodes of up to 8 K (8 steps of 2K each around the mean temperature of 40°C).[32] Thereby, an average thermovoltage was calculated from 10 measured values per temperature difference. The Seebeck coefficient was determined from the 8 averaged values for the thermovoltage using linear regression. The ends of the samples (thickness 0.3 mm) were painted with conductive silver to reduce the contact resistances. The electrical resistivity was measured using a 4-point-arrangement. A Keithley multimeter DMM2001 was used to measure the thermovoltage as well as the electrical resistance. All reported values represent mean values of 3-5 measurements on two strips cut from different positions of the compression molded plates. To quantify the TE performance, the Seebeck coefficient S was calculated from the measured TE voltage U divided by the applied temperature difference dT (S=U/dT). The power factor (PF) was determined by the equation PF = $S^2$ σ, whereas σ is the electrical conductivity, the reciprocal of the electrical resistivity.[33]

To measure the Hall coefficient, the PC based film samples were cut in a Hallbar shape (see Figure S2 and Table S1 for details) and characterized in a physical property measurement system (PPMS DynaCool from Quantum Design). The Hall resistance was measured in dependency of the applied magnetic field (Figure S3). The Hall coefficient, $A_h$, was obtained from the slope, $m$, of the measurement data and the measured thickness, $d$, of the sample ($A_h = m\,d$). The charge carrier concentration $n$ and mobility $\mu$ were calculated by $n = A_h\,q$ and $\mu = \sigma A_h$, respectively. Error propagation was applied to obtain the inaccuracies of $n$ and $\mu$, using the standard errors of the fit and the measurement errors of the used measurement devices.

TAS measurements on the PC based samples were performed on a Newport (TAS-1) transient absorption spectrometer, which is depicted schematically in Figure S4, while explained in detail elsewhere.[24,26,28] The employed light source was a Yb:KGW pulsed laser with the central wavelength at 1026 nm, a pulse duration of 170 fs, and a repetition rate of 1 kHz. The detection range was set between 550 and 910 nm, while a pump fluence of 15 mJ/cm$^2$ was used. The TE polymer composite samples were studied in the shape of the compression molded thin films with a thickness of 10 µm at room temperature. Several spots per sample were tested having similar signals (transmitted light). Areas with considerably less light transmission, indicating possible agglomerates, were avoided. Notably, PEEK based composite films could not be studied by TAS due to missing translucency.

### 3. Results and discussion

#### 3.1 Microscopical characterization

Figures 1a-c present typical SEM images of single-filler polycarbonate (PC) composite films upon incorporation of 1 wt% of SWCNT Tuball, MWCNT NC7000, and MWCNT CNS-

PEG, respectively. Inspection of these Figures reveals a relatively uniform dispersion and distribution of the CNT fillers within the polymer host as well as a good adhesion between CNTs and matrix. The same is the case for the hybrid-filler composites, in which two types of CNT fillers are embedded in the polymer matrix. Figure 1d shows a micrograph of a typical dispersion of 0.5% Tuball and 0.5% NC7000 fillers in PC, whereas Figure 1e depicts the corresponding image of a PC composite containing 0.5% Tuball and 0.5% CNS-PEG. The SEM image of the composite of PEEK with 1% Tuball given in Figure 1f also shows a suitable distribution and dispersion. Thus, the developed melt-mixing protocols allow the formation of uniform CNT networks in the PC and PEEK composites.

To investigate the macrodispersion of the CNTs, transmission light microscope was applied on thin sections made from the composite strands. Light microscopy images for the PEEK composites, each with 1 wt% of one CNT type are presented in Figure 2. While no remaining CNT agglomerates were found for the MWCNTs (NC7000, CNS-PEG), the composite with SWCNT Tuball shows some large remaining agglomerates which follow the typical shape of this kind of CNTs. The same findings concerning the macrodispersion of the three kinds of CNTs have already been reported for PC-based composites (compare Figure 5 in Ref. 31). This means that for the composites containing SWCNT Tuball, only the measurements which are based on small volumes of the film samples, such as Hall measurements, could be affected by these individual residual agglomerates.

**3.2 Thermoelectric performance and Hall measurements**

Figures 3a and b present the variation of the Seebeck coefficient (S) and electrical conductivity ($\sigma$) with CNT concentration in the two sets of single-filler PC and PEEK composite films, respectively, while all TE properties including the power factor PF are summarized in Table 1. Table 2 lists the corresponding TE parameters of the hybrid-filler PC and PEEK composites where more than one type of filler is incorporated. For the single-filler composites, in general, $\sigma$ increases in a nearly linear way with the loading of the CNTs (Table 1). In previous investigations, the percolation thresholds in PC composites were found to be about 0.05-0.075 wt% CNS-PEG, 0.1-0.2 wt% Tuball, and 0.25-0.5 wt% NC7000.[31] Percolation thresholds in PEEK composites were below 0.5 wt% CNS-PEG, at about 0.25 wt% Tuball, and between 1 and 3 wt% NC7000. Based on these thresholds, suitable CNT contents were incorporated in the polymer matrices.

In general, the Seebeck coefficient S exhibits a maximum at a certain CNT concentration and tends to decrease with further enhancement of the filler content. Depending on the CNT type, the determined S-values of the composites are very different. The highest S-values are obtained for SWCNT Tuball with 39.5 µV/K in PC (0.75 wt%) and 61.3 µV/K in PEEK composites (0.75 wt%). This correlates with the high S-value of 39.6 µV/K measured for the pure SWCNT powder.[10] Notably, a comparable high S-value was described for the same SWCNT type in poly(butylene terephthalate) (PBT) with 66.4 µV/K (at 5 wt%),[10] as well as in polypropylene with 47.9 µV/K (at 0.5 wt%).[9] Furthermore, depending on the filler loading a maximum of S is observed for both polymer matrices with SWCNT Tuball at 0.75 wt% addition. Comparable results with a maximum of S at around 0.5-1 wt% filler content were previously reported for PA 6,[10] ABS[10] or PP[9] based composites filled with SWCNT Tuball. For composites filled with MWCNT CNS-PEG, S-values of approx. 15.8 µV/K in PC and 13.4 - 16.1 µV/K in PEEK composites were determined. These Seebeck coefficients are thus somewhat higher than the S-value of the CNT powder (10.1 µV/K).[10] Only for PEEK/CNS-PEG composites, the maximum of S-values occurred at higher loading, namely at 1 wt%. In other polymer matrices, the incorporation of CNS-PEG led to S-values of 15.7 µV/K (PBT/2 wt%), 23.8 µV/K (PVDF/2 wt%), 17.5 µV/K (PP/2 wt%), 6.1 µV/K (ABS/4 wt%), 12.6 µV/K (PA6/5 wt%) or 4.2 µV/K (PA66/2 wt%).[10] For the PC and PEEK composites containing MWCNT NC7000, the highest electrical percolation threshold was determined among the three

CNT types. Therefore, only two composites could be measured which showed the lowest S-value (around 7 µV/K) compared to the composites with the other two CNT types. The comparatively low S-value correlates with the low S-value of the CNT powder itself, which was determined to be 6.3 µV/K.[10] In other polymer matrices, the incorporation of NC7000 led to S-values of 6.8 µV/K (PBT/2 wt%), 14.3 µV/K (PVDF/2 wt%), 9.5 µV/K (PP/2 wt%), 3.6 µV/K (ABS/4 wt%), 6.1 µV/K (PA6/5 wt%) or 6.3 µV/K (PA66/2 wt%).[10]

As expected, an increase in electrical conductivity with increasing CNT content was measured for all single-filler PC and PEEK composites (Table 1). Based on the general equation $\sigma = n\, e\, \mu$ (with e the electron charge), these findings are expected in terms of increasing $n$ when enhancing the number of CNTs within the composite network. Interestingly, the application of MWCNT CNS-PEG led to the significantly highest $\sigma$ at the same CNT concentration which may be related to its branched structure enabling a more dense and better developed CNT network. This highest $\sigma$ is in accordance with Hall measurements performed on PC based composites with 1 wt% of the three studied fillers as shown in Table 3. In particular, the PC-1 wt% CNS-PEG composite exhibits the highest σ among the three PC-based single-filler composites (Table 1). It also shows the highest $n$ which is about 17 times more when compared to that of the PC composite with 1 wt% NC7000, and 30 times more when compared to that of the composite with SWCNT Tuball. Also, µ is highest in the composite with CNS-PEG and lowest with Tuball. Consequently, this results in the lowest Hall coefficient $A_h$ for the composite with CNS-PEG followed by those for NC 7000 and Tuball based composites.

For the power factor, PF, as a measure of TE performance, the balance between the enhanced $\sigma$ and the decreasing S (above the named maxima) is important. In particular, $\sigma$ increases with increasing $n$ and $\mu$, while the S-value decreases with $n$. Thus, the highest PF value is not necessarily obtained for the composite with the highest S. Instead, PF depends on S² so that typically the change of S has a larger effect on the PF than a change of $\sigma$ which has a linear dependency on filler content. However, we see here the dominating effect of $\sigma$ on the PF due to the large increase of $\sigma$ (more than 2000 times when increasing NC 7000 concentration from 0.5 to 2 wt%, and 9 times for CNS-PEG) while S only increases maximal by 3%. Among all PC composites, the highest PF was found for PC/ 2 wt% CNS-PEG at $1.8 \cdot 10^{-2}$ µW/(m·K²), while having the highest $\sigma$ of all studied PC composites. On the contrary, only $1.4 \cdot 10^{-3}$ µW/(m·K²) was achieved for PC/ 0.75 wt% Tuball having the highest S-value of all studied PC composites or $6.1 \cdot 10^{-4}$ µW/(m·K²) for PC/ 2 wt% NC7000.

Interestingly, the trends described for PC composites are the same for PEEK composites (Figure 3b, Table 1). The Seebeck coefficient of the composites correlates with the S-value of the different CNT powders. Thus, the highest S-values are found for the composites with Tuball and the lowest for PEEK/NC7000 composites. Here, too, σ increases with the CNT content, with at comparable CNT contents the highest values being measured for PEEK/CNS-PEG composites. Furthermore, the PF also increases here with increasing CNT content and the relatively smaller S-value variations have a subordinate importance for this. It can be concluded that the exact values of the TE parameters Seebeck coefficient, volume conductivity and power factor are influenced by the type of matrix polymer, but the general trends concerning the differentiation between the different CNT types are the same. The branched MWCNTs result in the highest PF values of its composites, mainly due to the highest σ among all samples.

On an attempt to further improve the TE parameters, PC and PEEK were melt-mixed with a mixture of two types of CNT fillers. Figure 4 and Table 2 show TE data for hybrid-filler PC and PEEK composites when combining the MWCNT CNS-PEG (the filler that leads to the highest σ) with SWCNTs Tuball (leading to the highest S). In addition, MWCNT NC-7000 were combined with SWCNT Tuball. Thereby, the Tuball content was fixed to 0.5 or 0.75 wt% and different CNT contents of NC7000 or CNS-PEG were added. Again, it was found that with the increase of the total CNT content σ increases and S decreases. The increase in $\sigma$ is an expected consequence of the higher charge carrier concentration when more conductive filler is present. As expected, the addition of CNS-PEG leads to higher conductivity due to the higher conductivity of this filler, compared to the incorporation of NC7000. It is noticeable that, at the same (total) filler content, $\sigma$ of the PC based composites with hybrid filler systems is lower compared to the composites with a single CNT type. For example, a conductivity of 0.7 S/m was measured for PC/0.5% Tuball+0.5% NC7000, which is lower than 1.0 S/m and 1.1 S/m for PC/1% Tuball and PC/1% NC7000, respectively. With higher contents of the second filler (CNS-PEG, NC7000), the conductivity values approach those of the single-filler composite. Moreover, for PC/0.5 wt%Tuball+2 wt% CNS-PEG 42.3 S/m was measured, while PC/2 wt% CNS-PEG reached a conductivity of 72.8 S/m. Based on this, it seems that the hybrid CNT network in PC composites transports the charge carriers more poorly.

To investigate this hypothesis, the Hall measurements were extended to the sample PC/ 0.5 wt%Tuball+0.5 wt% CNS-PEG (Table 3). Interestingly, the μ value for the hybrid-filler system (0.19 cm$^2$ V$^{-1}$ s$^{-1}$) is significantly lower than for the single-filler systems (0.35-0.45 cm$^2$ V$^{-1}$ s$^{-1}$). This explains the significant drop of σ from 25.4 to 3.4 S/m when half of Tuball filler is replaced by CNS-PEG (Tables 1 and 2). Also, *n* correlates well with $\sigma$ of the composites. For the hybrid composite, PC/ 0.5 wt% Tuball+0.5 wt% CNS-PEG, *n* of 13.2·10$^{17}$ cm$^{-3}$ is measured, which is between the values of 1 wt% CNS-PEG on the one hand side and 1 wt% Tuball and NC7000 on the other side. Comparing $\sigma$, the same order can be found. In summary, the Hall measurements confirm that the charge carrier transport in the hybrid network is significantly slower, which is detrimental to achieve high $\sigma$ in such composites.

In contrast, at the same total filler concentration, $\sigma$ values of PEEK composites with hybrid filler systems are higher compared to the composites with only one CNT type. For example, $\sigma$ of PEEK filled with 0.5 or 0.75 wt% Tuball and 2 wt% CNS-PEG is with 128.5

S/m or 138.9 S/m, respectively, higher than PEEK/ 2-3 wt% CNS-PEG (51.9-97.1 S/m) and PEEK/ 0.5-0.75 wt% Tuball (around 2 S/m). Also, if 2 wt% NC7000 were added to PEEK filled with 0.5 or 0.75 wt% Tuball $\sigma$ increases to 18.5 or 35.7 S/m, whereby PEEK/ 2 wt% NC7000 was not electrically conductive. The reason for the different $\sigma$ behavior of the mixed filler systems in PC and PEEK could be the partial crystalline character of the PEEK. Whereas CNTs can act nucleating crystalline regions, they tend to be localized after the crystallization step in the amorphous regions of semi-crystalline polymers.[34-37] Thus, there is a concentration effect of the conductive fillers in the continuous amorphous PEEK phase, which has a positive effect on the common conductive network and can lower the electrical percolation threshold as the crystallinity within the PEEK composites increases. The hybrid network with its different morphology may influence the crystallinity in a different way than a single-filler network.

Meanwhile, it is found for both polymer types that the introduction of the second filler leads to a decrease of the S-value compared to the composites with Tuball only at the same total loading. In particular, the S of 1wt% single-filler Tuball PC composite drops from 36.7 µV/K to 28.2 µV/K and 22.6 µV/K, once 0.5wt% of Tuball is replaced by NC7000 or CNS-PEG, respectively (Table 2). In PEEK composites S reduces from 48.0 µV/K to 36.3 µV/K when comparing 1 wt% Tuball versus 0.5 wt% Tuball plus 0.5 wt% CNS-PEG. This reflects the variation of S of the different CNT types. This finding is related to the general tendency that S-values decrease with increasing filler content due to higher $n$ and $\sigma$ values.

Bearing in mind that the overall TE performance is expressed in terms of a PF given by the equation PF = $S^2 \sigma$, the results for single filler composites showed that the more prominent changes in $\sigma$ seem to play the dominant role over S. Thus, the replacement of a part of SWCNT Tuball by the second filler NC7000 or CNS-PEG has only marginal influence on the PF of the developed PC composites with 1 wt% total filler content. The PF of PC/1 wt% Tuball at $1.3 \cdot 10^{-3}$ µW/(m·K$^2$) could only increase up to $1.7 \cdot 10^{-3}$ µW/(m·K$^2$) if 0.5 wt% CNS-PEG was incorporated. The highest PF value of CNS-PEG filled composites is $1.8 \cdot 10^{-2}$ $\mu W/(m \cdot K^2)$ at 2 wt% loading, whereas the highest value in the hybrid-filler system is $1.5 \cdot 10^{-2}$ $\mu W/(m \cdot K^2)$ when combining 2 wt% CNS-PEG with 0.75 wt% Tuball. In PEEK composites, the addition of a second CNT type leads to an increase of PF. In general, the PF values of hybrid filler PEEK composites are in the range of $(1.1$ and $4.0) \cdot 10^{-2}$ $\mu W/(m \cdot K^2)$, whereas the PF of single CNT composites is one or two decades lower. For example, the PF of PEEK/0.5 wt% Tuball+0.5 wt% CNS-PEG at $3.8 \cdot 10^{-2}$ $\mu W/(m \cdot K^2)$ is higher compared to $7.2 \cdot 10^{-3}$ $\mu W/(m \cdot K^2)$ of PEEK/1 wt% Tuball or $4.8 \cdot 10^{-3}$ $\mu W/(m \cdot K^2)$ of PEEK/1 wt% CNS-PEG. The addition of 2 wt% NC7000 to 0.5-0.75 wt% Tuball increases PF from $7 \cdot 10^{-3}$ $\mu W/(m \cdot K^2)$ for PEEK/0.5-1 wt% Tuball to $(1.1-3.6) \cdot 10^{-2}$ $\mu W/(m \cdot K^2)$ for the hybrid filler composite. The main reason for this significant increase in PF is the above described $\sigma$ enhancement in these hybrid filler composites compared to the single filler composites.

Overall, the discussion of TE properties above has shown that the relationships concerning $\sigma$ can be well explained by the Hall measurements. However, the trends of S do not follow the expected anti-proportional relationships to $\sigma$. This makes the study of carrier dynamics on such composites particularly important to elucidate the mechanisms of free carrier conduction in the fabricated TE composite films, which was performed on PC based single- and hybrid-filler composites.

**3.3 Transient absorption spectroscopy (TAS) and exciton dynamics**

Figures 5a-c present typical spectra of the difference in optical density (*ΔOD*) as a function of wavelength at various delay times of the single-filler PC composites with 0.5 wt%, 0.75 wt%, and 1 wt% SWCNT Tuball. In agreement to previous TAS studies, a ground state negative *ΔOD* photobleaching profile is obtained at the vicinity of 730 nm for all studied films.[19-24] The observed *ΔOD* feature attenuates with time until it diminishes completely within several ps. The employed pump fluence of 15 mJ/cm$^2$ results to a good signal-to-noise ratio while ensuring the absence of unwanted heating and subsequent ablation effects on the composite films during the measurement. Notably, PC/NC7000 and PC/CNS-PEG composites exhibit similar $\varDelta OD$ profiles, whereas the pristine PC polymer exhibits no photobleaching signal under equivalent excitation conditions. This observation verifies that the observed negative $\varDelta OD$ profiles of the CNT-filled PC composites arise from the relaxation dynamics of the charged excitons within the incorporated CNTs.

The corresponding transient exciton decay dynamics of the PC/Tuball composites are shown in Figure 5d. The obtained optical density decay reflects the exciton lifetimes, indicative of the time period during which the free charge carriers (electrons and holes) are separated, and thus readily available for diffusion and charge transfer within the PC composites.[19-24] In particular, the lifetime component ($\tau_1$) is determined by means of a double-exponential fitting based on the equation y = y$_o$ + A$_1$ exp(-x/$\tau_1$) + A$_2$ (exp-x/ $\tau_2$). Table 1 lists the obtained lifetimes for the studied single-filler PC composites, while the exponential fitting procedure and the corresponding analysis is thoroughly described in the SI (Figure S5). Inspection of Figure 5d and Table 1 reveals that the PC/Tuball composites exhibit almost similar lifetimes ranging from 2 to 2.3 ps, while also having equivalent S. Moreover, the data of Table 1 and Figure 3a indicate similar findings for the other two single-filler composites, as no S improvement is observed upon increasing the concentration of NC7000 and CNS-PEG within the PC matrix. Notably, the latter composites exhibit similar exciton lifetimes, as was the case for the PC/Tuball composite films (Table 1). Thus, in the studied single-filler PC composites, where S appeared to be nearly independent of the CNT concentration within the PC matrix, the corresponding exciton lifetimes are found similar and more or less also independent of the filler concentration. Strikingly enough, on the contrary to the non-variation of S, the σ improved upon increasing the filler content within the PC-based single-filler composites. Indicatively, for the NC7000 composites σ improved from 0.04 to 8.5 S/m when the filler content increases from 0.5 wt% to 2 wt%, whereas the corresponding improvement for the CNS-PEG composites was from 7.8 to 72.8 S/m (Table 1 and Figure 3a).

We consider now effect of the filler type on the TE parameters and exciton lifetimes of the single-filler PC-based films. Figures 6a and b present the $\Delta OD$ profiles of 1 wt% PC/ 1wt% NC7000 and PC/ 1wt% CNS-PEG composites, respectively, while Figure 6c presents the corresponding exciton dynamics of the three 1wt% single-filler composites. As listed in Table 1, the obtained lifetimes for the PC/Tuball, PC/NC7000, and PC/CNS-PEG composite polymers are 2, 0.8, and 1.2, respectively, whereas the S values are 36.7, 9.2, and 15.8 µV/K. Remarkably, a direct correlation between the trends of S and exciton lifetimes emerges, as higher S values are obtained for longer exciton lifetimes (Table 1).

Nevertheless, even more interesting results emerge from the TAS studies of the hybrid-filler PC-based composites. Figures 7a and b show indicative examples of $\Delta OD$ profiles for PC with the hybrid-filler systems 0.5 wt% Tuball+0.5 wt% CNS-PEG and 0.5 wt% Tuball+2 wt% CNS-PEG composites. Similar to the single-filler composites, the TAS profile of the studied mixed-filler composites is dominated by a negative photobleaching feature that diminishes with time. However, in the case of hybrid-filler composites the $\Delta OD$ profile shows a wider bandwidth when compared to the corresponding profile of single-filler composites (Figure 5). This finding is rationalized due to the fact, that the incorporation of the additional filler within the PC polymer introduces more excitonic states for the electrons with different energy bandgaps.[19-24] The corresponding exciton dynamics are presented in Figure 7c, while probed at the maximum of the $\Delta OD$ profile for each time delay. Figure 7d depicts the exciton dynamics of the single-filler PC/1wt%Tuball composite along with these of hybrid-filler PC-system with 0.5 wt% Tuball+0.5 wt% CNS-PEG. Notably, the total content of filler within the latter two PC-based films is the same, i.e. 1 wt%. The obtained exciton lifetimes of all studied hybrid-filler composites are reported in Table 2. Inspection of exciton lifetimes and TE parameters of the PC-based Tuball+CNS-PEG system reveals a clear observation. Namely, as more MWCNT CNS-PEG filler is introduced within the composite containing 0.5 wt% Tuball, the more S decreased, while an improvement on the $\sigma$ is monitored. Remarkably, the same trends of S and σ are noted for the other three studied PC-based hybrid-filler systems, i.e. PC/0.75 wt% Tuball+CNS-PEG, PC/0.5 wt% Tuball+NC7000, and PC/0.75 wt% Tuball+NC7000 (Table 2).

Figure 8a presents the variation of S with respect to exciton lifetimes for both single-filler and hybrid-filler PC-based composites. An impressive linear correlation is noticed between S and the obtained exciton lifetimes, for both single-filler (blue line) and hybrid-filler (red line) composites. In fact, an almost linear correlation is obtained among the summary of data, independently of the CNT filler type and concentration within the host PC polymer matrix. Figure 8b shows the corresponding variation of $\sigma$ with exciton lifetimes. Strikingly, on the contrary to what is revealed for S, there is no correlation between σ and the obtained exciton lifetimes. Rather differently, σ appears to be totally independent from exciton lifetimes. These results provide new insights on the physical mechanisms of charge generation and transfer within the developed TE composite polymer films.

In principle, free charge carriers (electrons and holes) diffuse readily within a conductive material.[38] In a case where no external voltage or temperature variation occurs, the average carrier diffusion balances out and consequently no current is produced. Upon a temperature gradient however, at the hotter side of the TE composite there will be more variation in the energies of the free charge carriers when compared to the colder side.[25] In other words, at the hotter side more charge carriers are expected to be capable of occupying higher energetic states. At the same time, lower occupation for the lower energetic states is anticipated. Consequently, the free charge carriers of high energy states diffuse away towards the cold side of the TE composite, while the lower energy carriers move backwards to the hot side. In such scenario, a net current and thermovoltage is produced when one of the diffusion pathways is more profound than the other. We recall now on the definition of S for the TE composites, which is given by the formula S=U/dT, where U is the measured thermovoltage at a temperature difference dT. Based on the transient exciton dynamics of the studied composites it becomes apparent that in both single-filler and hybrid-filler composites the obtained longer lifetimes of the charged excitons favor the filling of the higher energy states. This increases the entropy between the two diffusion pathways of electrons from the hot side to the cold and backwards, and a higher thermovoltage is produced at the same temperature variation. Thus, it is revealed that for optimizing the S of TE composites, the key parameter is to achieve longer exciton lifetimes, i.e. slower recombination times between electrons and holes., upon varying the type and concentration of the CNT filler within the host polymer. Rather differently, as discussed in the previous section, the conductivity of both single-filler and hybrid-filler composites appears to be independent of the obtained exciton lifetimes (Figure 8b).

## 4. Conclusions

In conclusion, a variety of single-filler and hybrid-filler polymer composite films were prepared by melt-mixing aiming to investigate the filler effect on the TE performance. Polycarbonate (PC) and polyether ether ketone (PEEK) have been employed as polymer matrices. It can be concluded that the TE parameters (Seebeck coefficient S, conductivity $\sigma$ and power factor PF) of single- and hybrid-filler systems of both composites are mainly dependent on CNT type. The type of polymer influences the magnitude of the S-value, but the general tendencies that are CNT type dependent remain the same. The polymer type, especially whether amorphous (such as PC) or semi-crystalline (such as PEEK), plays a role in the formation of the electrically conductive network and thus also influences PF. Moreover, in order to shed light on the charge carrier generation and transfer mechanisms within the developed TE films, the PC-based polymers were studied my means of ultrafast laser time-resolved transient absorption spectroscopy (TAS). The systematic study of exciton dynamics, lifetime, and free charge carrier recombination, reveals novel insights on how the TE parameters are affected with respect to the employed CNT fillers. In brief, it is revealed that the TE parameter Seebeck coefficient S correlates directly with the obtained charged exciton lifetimes. In fact, the S-values of both PC-based single-filler and hybrid-filler composites exhibit an almost linear dependence with the exciton lifetime, with improved S values obtained for longer exciton lifetimes. This is rationalized due to the fact that the slower recombination times, favors the occupation of higher energy states by the free electrons, and thus increasing the thermovoltage potential between the hot and cold sides of the TE film. Rather differently, the second parameter of the TE, the electrical conductivity $\sigma$, is found to correlate mainly with the electron mobility within the composite films and not with exciton lifetimes. Interestingly enough, it was revealed that $\sigma$ decreases upon the replacement of a part of filler by a second filler within the PC composite, while maintaining the total filler content the same. The findings of the present study pave the way towards the design of advanced TE polymer composite architectures, while demonstrating that TAS is a performance technique for probing the carrier dynamics within TE energy harvesting materials.

**Supporting information.** Raman spectra of the employed CNTs. Hall measurements description. Transient Absorption Spectroscopy (TAS) experimental details and fitting procedures.

**Acknowledgements**

This work is funded within the frame of Horizon 2020 European Project "Innovative polymer-based composite systems for high-efficient energy scavenging and storage - InComEss", G.A. 862597.

**Figures and captions:**

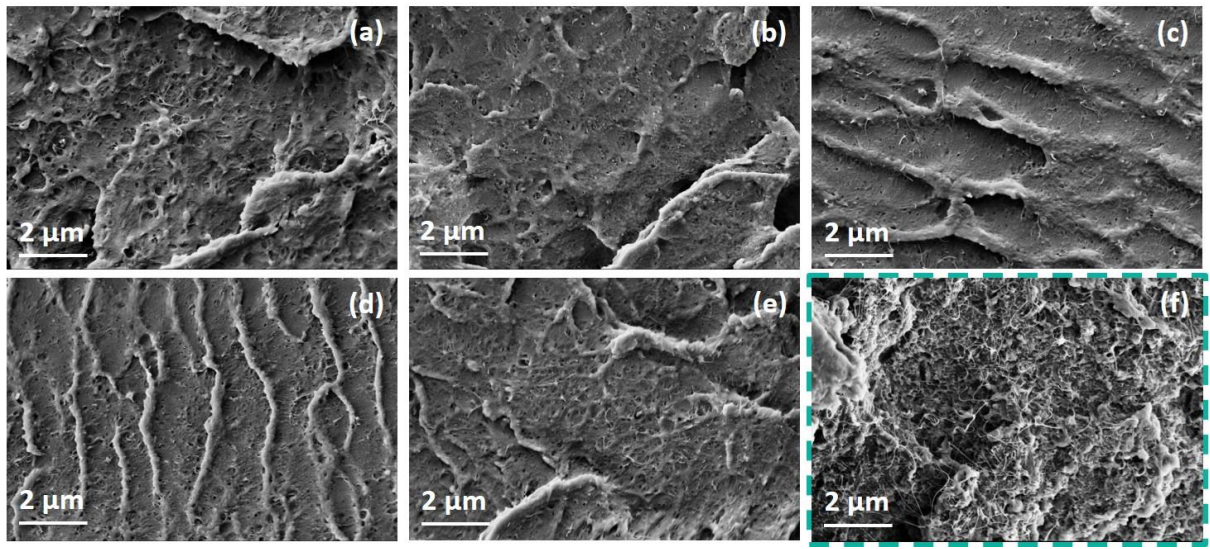

**Figure 1:** Scanning electron microscopy (SEM) images of 1 wt% single fillers in PC: **(a)** SWCNT Tuball, **(b)** MWCNT NC7000, and **(c)** MWCNT CNS-PEG. SEM images of hybrid fillers in PC: **(d)** 0.5 wt% Tuball - 0.5 wt% NC7000, and **(e)** 0.5 wt% Tuball - 0.5 wt% CNS-PEG. **(f)** SEM image of PEEK filled with 1 wt% Tuball.

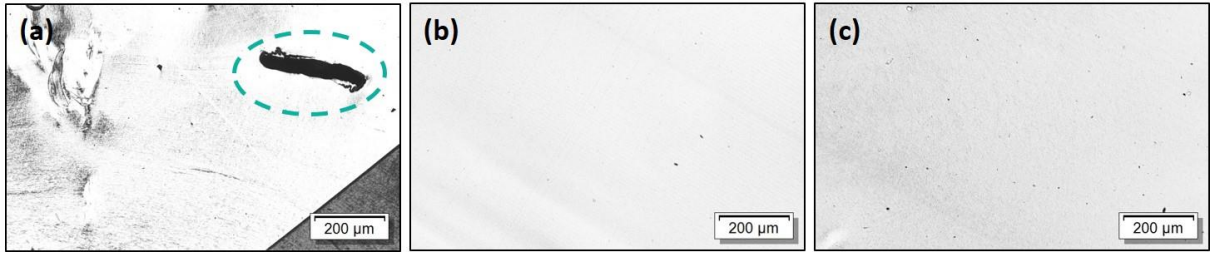

**Figure 2:** Transmission light microscopy (TLM) photos of 1 wt% single fillers in PEEK **(a)** SWCNT Tuball, **(b)** MWCNT NC7000, and **(c)** MWCNT CNS-PEG. The dashed circle marks a remaining SWCNT agglomerate.

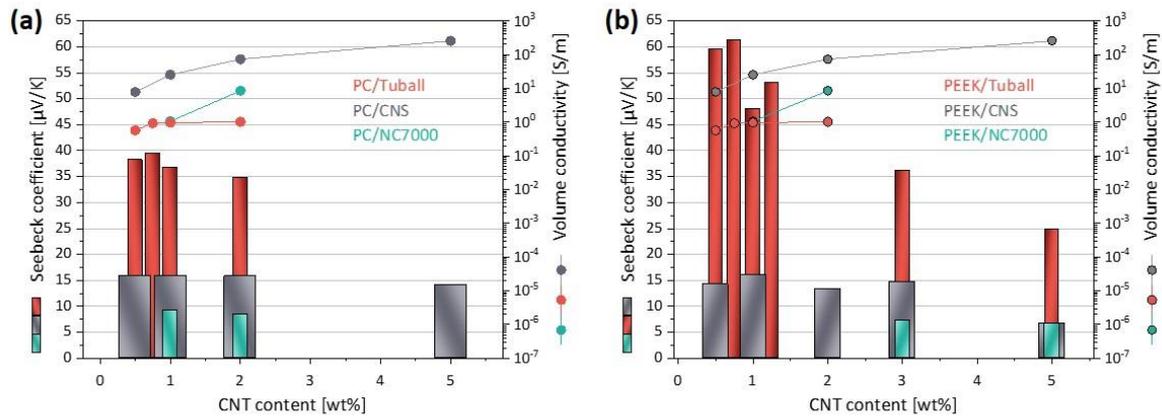

**Figure 3:** Seebeck coefficient (S) and volume conductivity ($\sigma$) of single-filler composites based on **(a)** polycarbonate, and **(b)** polyether ether ketone. The fillers are SWCNT Tuball, MWCNT NC7000, and CNS-PEG (CNS).

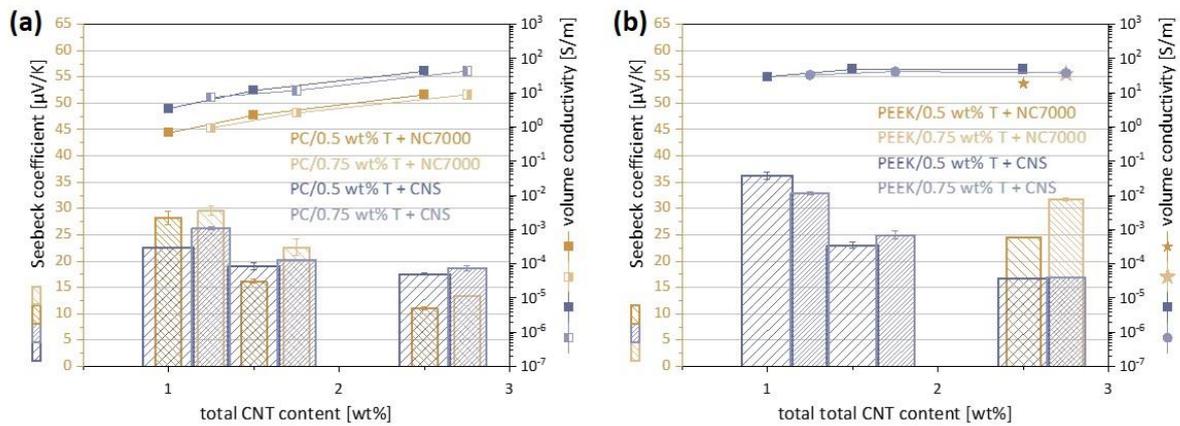

**Figure 4:** Seebeck coefficient (S) and volume conductivity ($\sigma$) of composites with hybrid fillers in **(a)** polycarbonate matrix, and **(b)** polyether ether ketone matrix. The hybrid fillers consist of 0.5 wt% or 0.75 wt% of SWCNT Tuball (T) plus different amounts of NC7000 or CNS-PEG (CNS). The TE properties are presented versus the total CNT content.

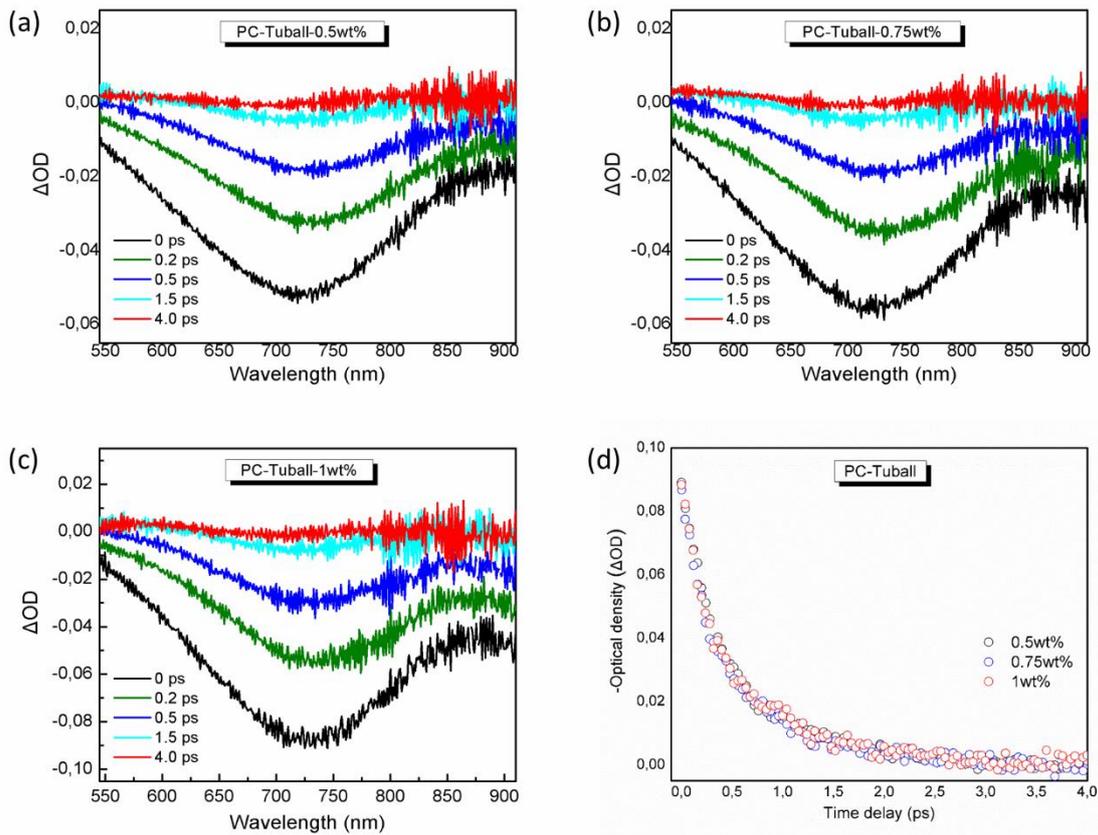

**Figure 5:** Transient absorption spectroscopy (TAS) spectra of the difference in optical density (*ΔOD*) as a function of wavelength for various delay times of PC-Tuball composites: **(a)** 0.5 wt%, **(b)** 0.75 wt%, and **(c)** 1 wt%. **(d)** Transient carrier dynamics of the single-filler PC-Tuball composites with different SWCNT loadings, probed at 730 nm, upon photoexcitation with a pump fluence of 15 mJ/cm$^2$.

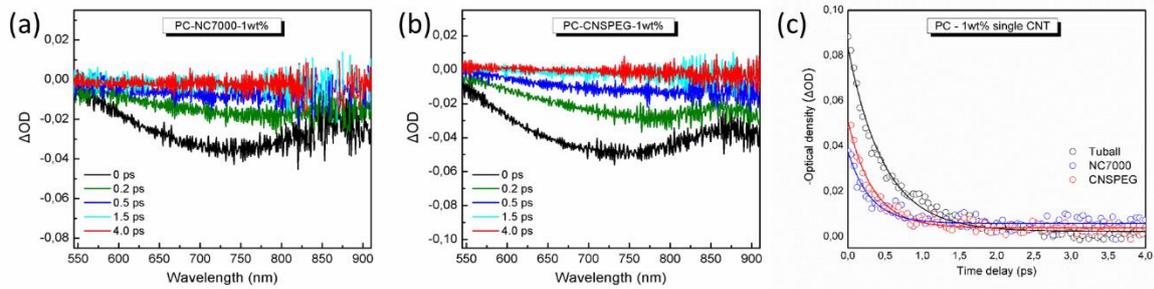

**Figure 6:** Transient absorption spectroscopy (TAS) spectra of the difference in optical density (*ΔOD*) as a function of wavelength for various delay times of PC based single-filler composites with **(a)** 1 wt% NC7000, and **(b)** 1 wt% CNS-PEG. **(c)** Transient carrier dynamics of the three single filler PC-1 wt% CNTs composites, probed at 730 nm, upon photoexcitation with a pump fluence of 15 mJ/cm$^2$. Empty symbols depict the experimental data, and solid lines represent the corresponding exponential fittings (see text).

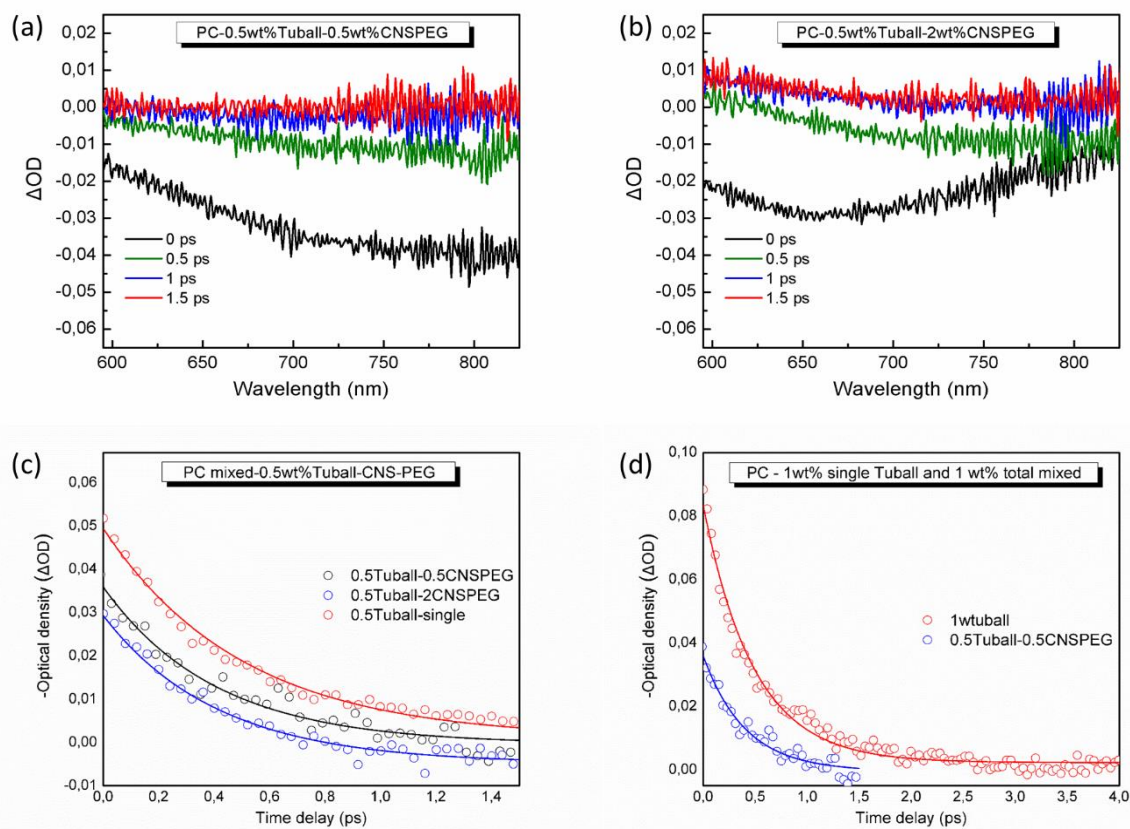

**Figure 7:** Transient absorption spectroscopy (TAS) spectra of the difference in optical density (*ΔOD*) as a function of wavelength for various delay times of mixed-filler composites: **(a)** PC-0.5 wt% Tuball-0.5 wt% CNS-PEG, and **(b)** PC-0.5 wt% Tuball-2 wt% CNS-PEG, **(c)** Transient carrier dynamics of the PC-0.5wt%Tuball-CNS-PEG polymers, probed at the maximum of ΔOD (see text), upon photoexcitation with a pump fluence of 15 mJ/cm$^2$. **(d)** Transient carrier dynamics of single-filler PC-1 wt% Tuball composite along with those of mixed-filler PC-0.5 wt% Tuball-0.5 wt% CNS-PEG. For **(c)** and **(d)** empty symbols depict the experimental data, and solid lines represent the corresponding exponential fittings (see text).

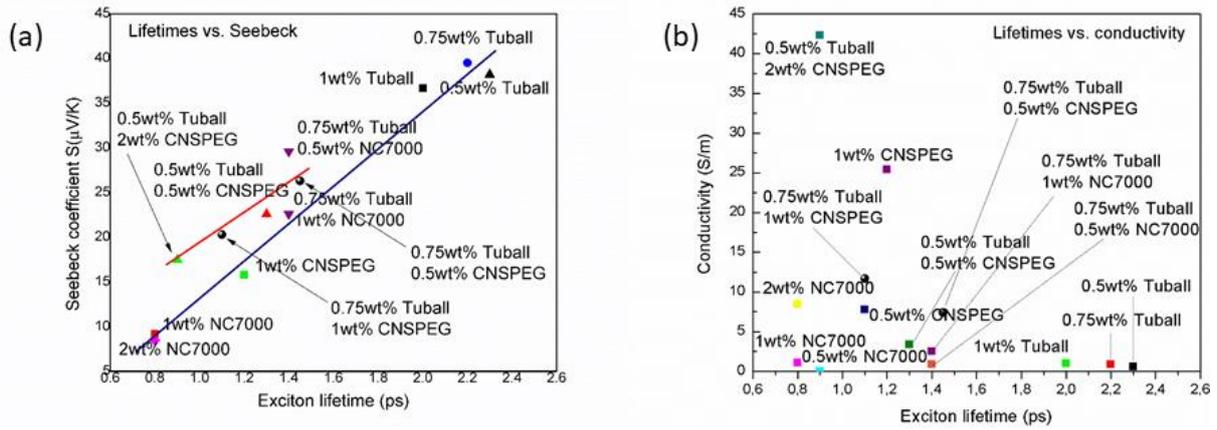

**Figure 8: (a)** Seebeck coefficient (S), and **(b)** volume conductivity (σ) variation with carrier lifetimes for single-filler and hybrid-filler polycarbonate (PC) composites. The blue and red lines are drawn as a guide to the eye for single-filler and hybrid-filler composites.

**Table 1:** TE properties of single-filler polycarbonate (PC) and polyether ether ketone (PEEK) composite polymer films, filled with Tuball, or NC7000, or CNS-PEG. Carrier lifetimes of the single-filler PC-based composites are also listed. *PEEK-based composites could not be measured.

| Sample | Volume conductivity (σ) [S/m] | Seebeck coefficient (S) [μV/K] | Power factor (PF) [μW/(m·K$^2$)] | Carrier lifetime [ps] (±0.1) |
|---|---|---|---|---|
| **PC-Tuball** | | | | |
| PC+0.5% Tuball | 0.6 | 38.2 ± 0.2 | 8.3×10$^{-04}$ | 2.3 |
| PC+0.75% Tuball | 0.9 | 39.5 ± 0.8 | 1.4×10$^{-03}$ | 2.2 |
| PC+1% Tuball | 1.0 | 36.7 ± 2.0 | 1.3×10$^{-03}$ | 2.0 |
| PC+2% Tuball | 1.0 | 37.8 ± 0.3 | 1.2×10$^{-03}$ | - |
| **PC-NC7000** | | | | |
| PC+0.5% NC7000 | 0.04 | - | - | 0.8 |
| PC+1% NC7000 | 1.1 | 9.2 ± 0.1 | 9.0×10$^{-05}$ | 0.8 |
| PC+2% NC7000 | 8.5 | 8.5 ± 0.0 | 6.1×10$^{-04}$ | 0.8 |
| **PC-CNS-PEG** | | | | |
| PC+0.5% CNSPEG | 7.8 | 15.9 ± 0.4 | 1.9×10$^{-03}$ | 1.1 |
| PC+1% CNSPEG | 25.4 | 15.8 ± 0.4 | 6.4×10$^{-03}$ | 1.2 |
| PC+2% CNSPEG | 72.8 | 15.8 ± 0.2 | 1.8×10$^{-02}$ | - |
| **PEEK-Tuball** | | | | |
| PEEK+0.5% Tuball | 2.2 | 59.4 ± 1.2 | 7.6×10$^{-3}$ | * |
| PEEK+0.75% Tuball | 1.8 | 61.3 ± 0.2 | 6.6×10$^{-3}$ | * |
| PEEK+1% Tuball | 6.2 | 48.0 ± 1.3 | 7.2×10$^{-3}$ | * |
| PEEK+1.25% Tuball | 7.2 | 53.0 ± 0.2 | 2.0×10$^{-2}$ | * |
| PEEK+3% Tuball | 21.8 | 36.2 ± 0.1 | 2.8×10$^{-2}$ | * |
| PEEK+5% Tuball | 65.0 | 24.8 ± 3.5 | 4.0×10$^{-2}$ | * |
| **PEEK-NC7000** | | | | |
| PEEK+1% NC7000 | - | - | - | * |
| PEEK+3% NC7000 | 5.2 | 7.3 ± 0.0 | 2.7×10$^{-4}$ | * |
| PEEK+5% NC7000 | 51.7 | 6.8 ± 0.0 | 2.35×10$^{-3}$ | * |
| **PEEK-CNS-PEG** | | | | |
| PEEK+0.5% CNSPEG | 8.8 | 14.2 ± 0.3 | 1.8×10$^{-3}$ | * |
| PEEK+1% CNSPEG | 18.4 | 16.1 ± 0.3 | 4.8×10$^{-3}$ | * |
| PEEK+2% CNSPEG | 51.9 | 13.4 ± 0.0 | 9.3×10$^{-3}$ | * |
| PEEK+3% CNSPEG | 97.1 | 14.7 ± 0.1 | 2.1×10$^{-2}$ | * |
| PEEK+5% CNSPEG | 51.7 | 6.8 ± 0.0 | 2.35×10$^{-3}$ | * |

**Table 2:** TE properties of hybrid-filler polycarbonate (PC) and polyether ether ketone (PEEK) polymer composite films, filled with Tuball plus NC7000 or CNS-PEG [a]

| Sample | Volume conductivity ($\sigma$) [S/m] | Seebeck coefficient (S) [µV/K] | Power factor (PF) [µW/(m·K$^2$)] | Carrier lifetime [ps] (±0.1) |
|---|---|---|---|---|
| **PC-0.5wt%Tuball-NC7000** | | | | |
| PC+0.5%Tuball+ 0.5% NC7000 | 0.7 | 28.2 ± 0.1 | 5.3x10$^{-04}$ | 1.6 |
| PC+0.5%Tuball +1%NC7000 | 2.2 | 16.2 ± 0.4 | 5.7x10$^{-04}$ | 1.0 |
| PC+0.5%Tuball+2%NC7000 | 8.8 | 11.0 ± 0.3 | 1.1x10$^{-03}$ | 0.9 |
| **PC-0.75wt%Tuball-NC7000** | | | | |
| PC+0.75%Tuball+ 0.5%NC7000 | 0.9 | 29.6 ± 0.9 | 8.2x10$^{-04}$ | 1.4 |
| PC+0.75%Tuball+ 1%NC7000 | 2.5 | 22.6 ± 0.2 | 1.3x10$^{-03}$ | 1.4 |
| PC+0.75%Tuball+ 2%NC7000 | 8.7 | 13.5 ± 0.1 | 1.6x10$^{-03}$ | - |
| **PC-0.5wt%Tuball-CNS-PEG** | | | | |
| PC+ 0.5%Tuball+ 0.5% CNSPEG | 3.4 | 22.6 ± 0.1 | 1.7x10$^{-03}$ | 1.3 |
| PC+ 0.5% Tuball+ 1% CNSPEG | 11.6 | 19.0 ± 0.6 | 4.3x10$^{-03}$ | 1.1 |
| PC+ 0.5% Tuball+ 2% CNSPEG | 42.3 | 17.5 ± 0.2 | 1.3x10$^{-02}$ | 0.9 |
| **PC-0.75wt%Tuball-CNS-PEG** | | | | |
| PC+ 0.75%Tuball+ 0.5% CNSPEG | 7.4 | 26.3 ± 0.3 | 5.1x10$^{-03}$ | 1.4 |
| PC+ 0.75%Tuball+ 1% CNSPEG | 11.7 | 20.3 ± 0.0 | 1.8x10$^{-03}$ | 1.1 |
| PC+ 0.75%Tuball+ 2% CNSPEG | 42.7 | 18.7 ± 0.4 | 1.5x10$^{-02}$ | 2.4 |
| **PEEK-0.5wt%Tuball-NC7000** | | | | |
| PEEK+0.5%Tuball +2%NC7000 | 18.5 | 24.5 ± 0.1 | 1.1x10$^{-2}$ | * |
| **PEEK-0.75wt%Tuball-NC7000** | | | | |
| PEEK + 0.75% Tuball+2% NC7000 | 35.7 | 31.8 ± 0.3 | 3.6x10$^{-2}$ | * |
| **PEEK-0.5wt%Tuball-CNS-PEG** | | | | |
| PEEK+0.5%Tuball +0.5%CNSPEG | 29.0 | 36.3 ± 0.7 | 3.8x10$^{-2}$ | * |
| PEEK+0.5%Tuball + 1%CNSPEG | 48.4 | 23.0 ± 0.6 | 2.6x10$^{-2}$ | * |
| PEEK+0.5%Tuball + 2%CNSPEG | 128.5 | 16.6 ± 0.1 | 3.6x10$^{-2}$ | * |

| PEEK-0.75wt%Tuball-CNS-PEG | | | | |
|---|---|---|---|---|
| PEEK+0.75%Tuball+ 0.5% CNSPEG | 33.0 | 32.9 ± 0.3 | 2.0x10⁻² | * |
| PEEK+0.75%Tuball+ 1% CNSPEG | 41.5 | 24.9 ± 0.8 | 2.7x10⁻² | * |
| PEEK+0.75%Tuball+ 2% CNSPEG | 138.9 | 17.0 ± 0.1 | 4.0x10⁻² | * |

*a* Carrier lifetimes of the mixed-filler PC-Tuball-based composites are also listed.
*PEEK-based composites could not be measured.

**Table 3:** Hall coefficient $A_h$, charge carrier concentration $n$ and mobility $\mu$ obtained by Hall measurements for selected PC based composites.

| Sample | $A_h$ (cm³ C⁻¹) | $n$ (10¹⁷ cm⁻³) | $\mu$ (cm² V⁻¹ s⁻¹) |
|---|---|---|---|
| PC+ 1 wt% NC7000 | 25.3 ± 2.2 | 2.5 ± 0.2 | 0.41 ± 0.06 |
| PC+ 1 wt% CNS-PEG | 1.4 ± 0.6 | 43.6 ± 20.5 | 0.45 ± 0.21 |
| PC+ 1 wt% Tuball | 43.6 ± 2.9 | 1.5 ± 0.1 | 0.35 ± 0.05 |
| PC+ 0.5 wt% Tuball + 0.5 wt% CNS-PEG | 4.7 ± 0.3 | 13.2 ± 0.1 | 0.19 ± 0.03 |

**ToC figure:**

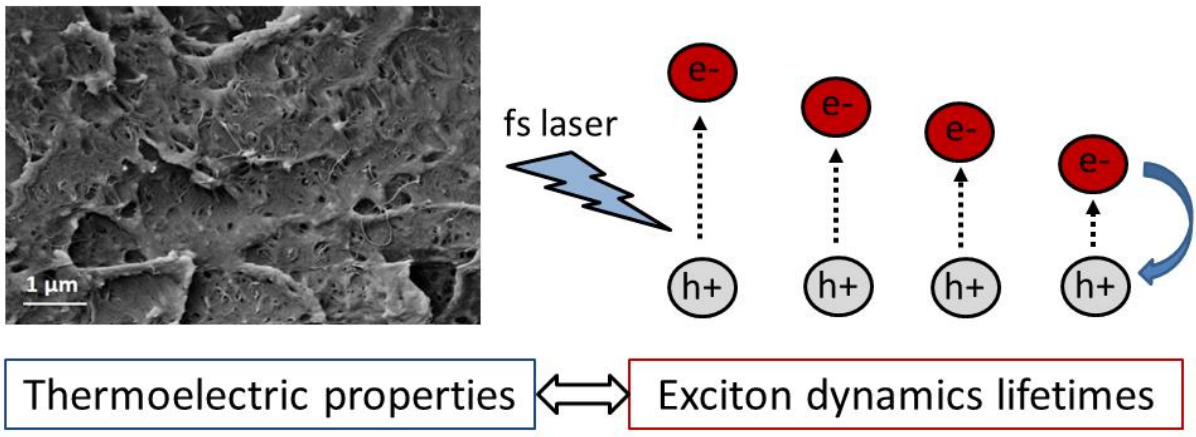

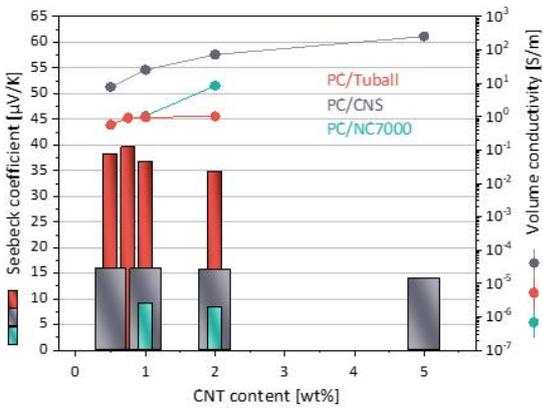
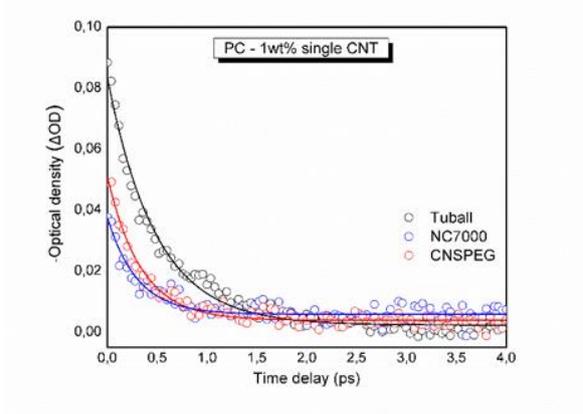